\documentclass[12pt]{JHEP3}
\usepackage{amsmath,amssymb,epsfig}

\newcommand{\ra}{\rightarrow}

\newcommand{\bc}{\begin{center}}
\newcommand{\ec}{\end{center}}
\newcommand{\ba}{\begin{array}}
\newcommand{\ea}{\end{array}}
\newcommand{\beq}{\begin{equation}}
\newcommand{\eeq}{\end{equation}}
\newcommand{\bea}{\begin{eqnarray}}
\newcommand{\eea}{\end{eqnarray}}
\newcommand{\bmx}{\begin{pmatrix}}
\newcommand{\emx}{\end{pmatrix}}
\newcommand{\nn}{\nonumber}
\newcommand{\al}{\alpha}
\newcommand{\be}{\beta}

\newcommand{\m}{\mu}
\newcommand{\n}{\nu}

\newcommand{\te}{\theta}

\newcommand{\G}{\Gamma}

\newcommand{\nt}{\noindent}
\newcommand{\del}{\partial}
\newcommand{\half}{\frac{1}{2}}
\newcommand{\tr}{{\rm tr}}
\newcommand{\tbar}{{\overline t}}

\newcommand{\eref}[1]{Eq.~(\ref{#1})}

\newcommand{\cE}{{\cal E}}
\newcommand{\cF}{{\cal F}}
\newcommand{\cO}{{\cal O}}
\newcommand{\cP}{{\cal P}}
\newcommand{\cR}{{\cal R}}
\newcommand{\cZ}{{\cal Z}}
\newcommand{\cU}{{\cal U}}
\newcommand{\cW}{{\cal W}}
\newcommand{\zbar}{{\bar z}}
\newcommand{\cT}{{\cal T}}

\def\IB{\relax{\rm I\kern-.18em B}}
\def\IC{{\relax\hbox{\kern.3em{\cmss I}$\kern-.4em{\rm C}$}}}
\def\ID{\relax{\rm I\kern-.18em D}}
\def\IE{\relax{\rm I\kern-.18em E}}
\def\IF{\relax{\rm I\kern-.18em F}}
\def\II{\relax{\rm I\kern-.18em I}}
\def\Id{\relax{1\kern-.32em 1}}
\def\IG{\relax\hbox{$\inbar\kern-.3em{\rm G}$}}
\def\IR{\relax{\rm I\kern-.18em R}}
\newcommand\sfrac[2]{{\textstyle\frac{#1}{#2}}}
\newcommand\ncr[2]{\begin{pmatrix}{#1}\\{#2}\end{pmatrix}}
\newcommand\shalf{{\textstyle\frac12}}

\normalsize

\title{Noncritical String Correlators, Finite-N Matrix Models and the
Vortex Condensate}
\author{Anindya Mukherjee\,\footnote{Email: anindya\_m@theory.tifr.res.in}\,
and Sunil Mukhi\,\footnote{Email: mukhi@tifr.res.in}\\
\it Tata Institute of Fundamental Research,\\
\it Homi Bhabha Rd, Mumbai 400 005, India}
\abstract{We carry out a systematic study of correlation functions of 
momentum modes in the Euclidean $c=1$ string, as a function of the
radius and to all orders in perturbation theory. We obtain simple
explicit expressions for several classes of correlators in terms of
special functions. The Normal Matrix Model is found to be a powerful
calculational tool that computes $c=1$ string correlators even at
finite $N$. This enables us to obtain a simple combinatoric formula
for the $2n$-point function of unit momentum modes, which after
T-duality determines the vortex condensate. We comment on possible
applications of our results to T-duality at $c=1$ and to the 2d black
hole/vortex condensate problem.}

\preprint{hep-th/0602119\\ TIFR/TH/06-03}


\begin{document}

\section{Introduction}
\label{Introduction}

The $c=1$ string (an excellent review is Ref.\cite{Klebanov:1991qa})
is a perturbatively consistent string theory in two spacetime
dimensions. One of its attractive features is that it is solvable:
from the powerful techniques of Matrix Quantum Mechanics (MQM),
correlation functions of the momentum modes (``tachyons'') can be
determined to all orders in the string coupling (inverse cosmological
constant). This holds true even in the Euclidean theory at finite
radius $R$. Another feature is that its generalisation to the type 0
noncritical string, has similar properties in perturbation theory but
is believed to also be non-perturbatively well-defined.

This makes the $c=1$ string and its cousins a good laboratory to study
various open questions in string theory. Two such questions that we
would like to understand better in the noncritical context are the
properties of string-scale black holes, and the nature of various
dualities, including open-closed string
duality\cite{McGreevy:2003kb}\cite{Martinec:2003ka}\cite{Klebanov:2003km}.
Much work has been done on the former (some interesting recent studies
can be found in
Refs.\cite{Karczmarek:2004bw}\cite{Kutasov:2005rr}\cite{Giveon:2005jv}),
while the latter question has also yielded some important
illuminations\cite{Sen:2003iv}\cite{Gaiotto:2003yb}%
\cite{Maldacena:2004sn}\cite{Hashimoto:2005bf}\cite{Gaiotto:2005gd}%
\cite{Mukherjee:2005aq}\cite{Ellwood:2005nt}.

It is known\cite{FZZconjecture}\cite{Kazakov:2000pm} that basic
properties of black holes in noncritical string theory are controlled
by condensates of winding tachyons in the Euclidean-continued
background. These are thermal tachyons: strings winding around the
compact time direction. It would therefore be useful to know the
correlators of winding modes in Euclidean noncritical string theory to
all orders in the string coupling (and even nonperturbatively in the
stable type-0 case) as a function of $\mu$ and $R$, where $\mu^{-1}$
is the inverse string coupling and $R$ is the radius of the Euclidean
direction (inverse temperature). From the matrix model point of view,
winding modes are related to the nonsinglet sector of the model, in
which the eigenvalue fermions are no longer free but mutually
coupled\cite{Gross:1990md}\cite{Boulatov:1991xz}. Computing
correlators in this way is a harder task\cite{Maldacena:2005hi} and
has raised some new puzzles involving leg factors which we will
discuss in a later section. But one way to find the desired
correlators is to assume that T-duality holds and perform it on the
momentum correlators. This provides one of our motivations to study
momentum correlators in the Euclidean theory in more explicit detail
than has already been done.

As mentioned above, momentum correlators in the Euclidean $c=1$ string
are known in principle. They are summarised in the Toda hierarchy or
$W_\infty$ symmetries\cite{Dijkgraaf:1992hk}, or Hirota bilinear
equations, or Normal Matrix Model (NMM)\cite{Alexandrov:2003qk}, all
of which are supposed to be mutually equivalent. For the special case
of self-dual radius $R=1$ of the Euclidean time direction, they are
encoded in a Kontsevich-Penner matrix
model\cite{Imbimbo:1995yv}\cite{Imbimbo:1995ns} (see also
\cite{Mukhi:2003sz}\cite{Aganagic:2003qj}). We will summarise some
relevant information about these solutions below. But while all these
formal solutions allow us to extract the perturbation series for any
specific correlator after a sufficient amount of work, we do not have
many explicit answers in terms of special functions depending on the
radius $R$ and inverse string coupling $\mu$.

At finite radius, correlators have been computed mostly at tree-level
(corresponding to the dispersionless limit of the Toda hierarchy) or
to a few low orders in perturbation theory. For example, while the
$2n$-point function of $n$ unit winding modes and $n$ anti-winding
modes is known as a function of $n$ and $R$ at tree level
\cite{Moore:1992ga}\cite{Kazakov:2000pm}, an explicit
expression for the same correlator to all orders in perturbation
theory does not seem to exist in the literature\footnote{A
differential equation for these correlators was written down
in\cite{Kazakov:2000pm} together with an iterative solution to a few
orders. Related work on Euclidean correlators can be found in
Refs.\cite{Alexandrov:2001cm},\cite{Alexandrov:2002fh}.}. To be more
specific, denote by $T_q$ the tachyons of momentum $q=n/R$, and by
$\cT_q$ the tachyons of $n$ units of winding, where $q=nR$ is the
value of $p_L=-p_R$ in vertex operator language. An explicit form is
known for $\langle (T_{-1/R})^n(T_{1/R})^n\rangle$ at tree level. The
T-dual of this expression was used in Ref.\cite{Kazakov:2000pm} to
extract the critical behaviour of the Sine-Liouville theory defined by
perturbing the original $c=1$ string with $\cT_{-R} + \cT_{R}$ and
then tuning the cosmological constant $\mu$ to zero. In particular,
Ref.\cite{Kazakov:2000pm} showed that a sensible theory exists after
this tuning, but only when the radius of the Euclidean direction lies
in the range $1<R<2$.

One would like to know the structure of this correlator to all string
loop orders. Accordingly, in what follows we will study $\langle
(T_{-1/R})^n(T_{1/R})^n\rangle$ in detail, and one of our main results
will be a simple formula for this correlator as a function of $\mu$
and $R$ for every $n$. We expect this to lead to a better
understanding of the exponentiated correlator $\langle
\exp(T_{-1/R}+ T_{1/R})\rangle$, which in turn is T-dual to the vortex
condensate $\langle \exp({\cal T}_{-R}+ {\cal T}_{R})\rangle$ that
relates directly to Euclidean 2d black holes.

Another motivation for our work is to understand T-duality of the
$c=1$ matrix quantum mechanics. This is established at the level of
spectrum of states, since the partition function without perturbations
is known to be T-dual\cite{Gross:1990ub}. Also, a formal argument has
been given\cite{Kazakov:2000pm} that the winding correlators, like the momentum
correlators, are given by a Toda
hierarchy\footnote{For a discussion of T-duality in type 0A,B
matrix models, see Ref.\cite{Yin:2003iv}.}. 
However, to our knowledge, beyond this
result and a computation in \cite{Maldacena:2005hi}, there has been no
direct comparison of correlators in the momentum and winding
sectors\footnote{In addition to pure momentum or pure winding
correlators, one would also like to know the correlators for a mixture
of momentum and winding modes. In this case one has no choice but to
tackle the difficult nonsinglet sector problem. The system is not
expected to be integrable and the correlation functions are not known
so far.}. A convincing test of T-duality would consist of computing
pure-momentum correlators in terms of free fermion eigenvalues,
T-dualising the answers and comparing them with pure-winding
correlators computed from the nonsinglet Hamiltonian. Ideally this
should even be done beyond tree level. Although we will not be able to
carry out such a test here, we have tried to systematise one side of
the duality in a way that can be eventually compared with the other
side when nonsinglet computations become more practicable.

In particular, the most direct way to check T-duality comes from
comparing two-point functions. Accordingly we work out all two-point
functions of momentum modes. In Ref.\cite{Maldacena:2005hi}, the
two-point function of unit-momentum modes was computed and an attempt
made to match the leading result with a computation in the first
nonsinglet sector of the matrix model, namely the adjoint sector. The
comparison revealed the presence of unexplained normalisation factors.
It was pointed out in Ref.\cite{Maldacena:2005hi} that if one could
compute two-point functions of more general winding modes, namely
$\langle {\cal T}_{-nR}{\cal T}_{nR}\rangle$, one might be able to
shed some light on these normalisation factors. With this motivation
we have performed this computation and obtained a simple explicit
result, again as a function of $\mu$ and $R$ and for all $n$. In a
later section we discuss the relation to the non-singlet sectors.

Our initial computations have been performed using both the MQM and a
model of constant matrices called the Normal Matrix Model
(NMM)\cite{Alexandrov:2003qk}, with perfect agreement between the
answers. In the former case we used the known infinite-radius
correlators in the physical MQM (real and noncompact
time)\cite{Moore:1991sf}, and a formula which converts these to the
correlators for Euclidean compact time\cite{Klebanov:1991ai}. In the
latter case, we will describe how one performs computations after
reducing the NMM to eigenvalues. One surprise emerging from comparison
of the two approaches is that the NMM successfully computes
correlators even when the matrices are of finite rank $N$, a stronger
property than was claimed in Ref.\cite{Alexandrov:2003qk}, who did
however suggest that the model contains some information even at
finite $N$. We find that it actually contains {\it complete}
information at finite $N$ in the following sense: given a correlator,
there is a minimum value $N_{min}$ such that this correlator when
computed in the NMM gives the correct result, to all orders in
$1/\mu^2$, for all $N>N_{min}$. This makes the NMM a potentially
powerful combinatoric tool. We then go on to demonstrate its power by
deriving a combinatoric formula for the general correlation function
$\langle (T_{-1/R})^n(T_{1/R})^n\rangle$ for any $n$.

We start in Section \ref{mqmnmmreview} by describing the two relevant
matrix models, Matrix Quantum Mechanics and Normal Matrix Model. The
former is too well-known to need a detailed discussion and we skip
directly to the calculational techniques and answers. For the latter,
we review the model in some detail, with special attention to the role
of the matrix rank $N$. In Section \ref{mqmcorr} we work out some
relevant correlators as a function of $\mu$ and $R$ from MQM. In
Section \ref{nmmcorr} we reproduce these correlators from the NMM,
where we note the phenomenon that for a fixed correlator, the NMM at
any $N$ greater than a minimum value gives the complete answer. After
a discussion of why this works, we use this property to derive a
combinatoric formula for correlators of any number of unit momentum
modes. In Section \ref{applications} we discuss applications of these
results to some physically interesting problems, and conclude in
Section \ref{conclusions}. Several calculational details are presented
in the appendices.

\section{Matrix Quantum Mechanics and Normal Matrix Model}
\label{mqmnmmreview}

\subsection{Matrix Quantum Mechanics}

Matrix Quantum Mechanics is a model of a single $N\times N$ hermitian
time-dependent matrix $M(t)$. In the absence of perturbations, the
partition function of the model is given by:
\beq
\cZ_{MQM}^{(N)} = \int [dM] \exp\left[-N \int dt\,\tr\left((D_t M)^2 +
M^2 \right)\right]
\eeq
where $D_tM\equiv \dot{M} + i[A_t,M]$ is the covariant derivative with
respect to the time component of a gauge field.

The gauge field acts as a Lagrange multiplier and projects the model
to the singlet sector, which is a system of $N$ non-interacting
non-relativistic fermions moving in an inverted harmonic oscillator
potential. In the double-scaling limit, the fermi sea is filled nearly
to the top and the number of fermions is taken to infinity. The scaled
distance to the top of the potential, $\mu$, is kept finite and
corresponds to the cosmological constant. This model provides a
description of 2D string theory, with $\mu^{-1}$ playing the role of
the string coupling $g_s$.

The physical modes of 2D string theory can be constructed in terms of
fermion eigenvalues. In \cite{Moore:1991sf} this model was used to
calculate correlation functions of $c=1$ string theory at infinite
radius. One starts by computing correlators of free-fermion bilinears,
which in turn can be used to extract correlators of the loop
operators:
\beq
{\cal O}(k,\ell) = \int dt\,e^{ikt}\,\tr\, e^{-\ell M(t)}
\eeq
Extracting the leading behaviour of these loops for small
$\ell$, one has
\beq
{\cal O}(k,\ell)\sim \ell^{|k|}\,T_k
\eeq
The $T_k$ are identified with the $c=1$ string theory tachyons. When
compared with the corresponding operators in Liouville theory, there
is a change of normalisation:
\beq
\label{mqmnorm}
T_k|_{MQM}=\G(|k|)~T_k|_{Liouville}
\eeq
However this fact will not be relevant for us, since in what follows we
will always work with the operators $T_k$ in the MQM basis, i.e. the
LHS of the above equation.

When the time direction is Euclidean and compact, we are in the finite
temperature theory. Starting from the infinite-radius correlator, one
can show\cite{Klebanov:1991ai} that correlators in the Euclidean
theory at finite radius are obtained as:
\beq
\langle T_{q_1}T_{q_2}\cdots T_{q_n} \rangle_R 
= \frac{\frac{1}{2R}\del_\m}{{\rm
sin}\left(\frac{1}{2R}\del_\m\right)}
\langle T_{q_1}T_{q_2}\cdots T_{q_n} \rangle_\infty
\label{finiteR}
\eeq
In addition one must replace the momentum-conserving $\delta$-function
as:
\beq
\label{deltafactor}
\delta\Big(\sum_i q_i\Big) \to R\, \delta_{\sum_i q_i,0}
\eeq

The above prescriptions follow from the fact that the compact radial
direction introduces an additional factor in the loop momentum
integrals of the infinite-radius calculation, and this factor can now
be taken out of the integrals whence it becomes a differential
operator acting on the infinite-radius answer\footnote{It is also
possible to calculate correlators directly at finite radius using the
``reflection coefficient'' formalism of
Ref.\cite{Dijkgraaf:1992hk}. Though we will not use this here, it
would be interesting to know if our explicit results follow as easily
in that approach.}.

In the finite-temperature theory, the above modes can be thought of as
carrying ``momentum'' in the time direction. In this situation one
also expects to find winding modes corresponding to the thermal
scalars of finite-temperature string theory. Many physical properties
of string theory are encoded in these degrees of freedom, which are
therefore quite important to study. To find them in the matrix model
we must go beyond the singlet sector, in which the gauge field is
topologically trivial and can be gauged away. Consider the
gauge-invariant Wilson-Polyakov loop variable:
\beq
W_\cR = \tr_{\cR}\, P\,\exp(i\oint A_t dt)
\eeq
where the trace is performed in the representation $\cR$ of
$SU(N)$. When $\cR$ is the fundamental representation, this is to be
associated with a unit winding mode:
\beq
\label{wilsonwinding}
\cW_{\cR=N}\sim \cT_R 
\eeq

Similarly the trace in the anti-fundamental will be $\cT_{-R}$. One
can also have loops where the trace continues to be in the fundamental
but the contour winds multiple times over the Euclidean time
direction. Computation of the correlation functions of all these
Wilson-Polyakov loops is done by observing that in their presence, the
matrix model receives contributions from definite non-singlet
sectors. In these sectors it reduces to eigenvalue fermions but now
with mutual interactions. For example, the two-point function of
unit winding modes can be identified as follows:
\beq
\langle \cT_{-R}\cT_R\rangle\Big|_{\hbox{\tiny\bf Liouville theory}} \sim 
\langle \cW_{\bar N}\cW_N\rangle\Big|_{\hbox{\tiny\bf MQM}}\sim 
\langle \cW_{\hbox{\tiny\bf adjoint}}\rangle\Big|_{\hbox{\tiny\bf MQM}}
\eeq
Thus computing the partition function of MQM in the adjoint sector
determines the two-point function of winding modes. Since in principle
this is an independent computation from that of the momentum tachyon
correlators, it can actually be used to check T-duality of the $c=1$
string. We will return to this issue in a subsequent section.

\subsection{Normal Matrix Model}

The Normal Matrix Model (NMM)\cite{Alexandrov:2003qk} is a relatively
simple model of a complex matrix $Z$ and its Hermitian adjoint, with
the constraint that the two commute (hence $Z$ is said to be
``normal''). The potential is polynomial with an additional
logarithmic piece. The matrix $Z$ is constant rather than
time-dependent, so in this sense it is more similar to the $c<1$
string backgrounds which do not have a time direction\footnote{Perhaps
this is the underlying reason why the NMM describes Euclidean $c=1$
strings at an arbitrary radius $R$, but does not have a simple
$R\to\infty$ limit where one might recover the Lorentzian theory.}.

The NMM is proposed to describe the correlators of the $c=1$ string to
all orders in perturbation theory, as follows. Let us introduce its
partition function:
\beq
\label {ZNMM}
{\cal Z}^{(N)}_{NMM}(\nu,t,\overline{t}) 
= \int [dZ dZ^\dag]\ e^{{\bf tr}\left( -\n
(ZZ^\dag)^R + \left(R\n - N + \frac{R-1}{2}\right){\rm\,log}\,ZZ^\dag - \n
\sum_{k=1}^{\infty}(t_k Z^k + \overline{t}_k {Z^\dag}^k)\right)}
\eeq
Here $R,\nu$ are some (in general, complex) parameters, which will
correspond to the compactification radius of Euclidean time and the
cosmological constant respectively. The parameters $t_k$,
$\overline{t}_k$ are couplings to the gauge-invariant operators $\tr
Z^k$, $\tr {Z^\dag}^k$ and $Z$, $Z^\dag$ are $N\times N$ matrices
satisfying:
\beq
[Z,Z^\dag]=0
\eeq
The operators $\tr Z^k,\tr {Z^\dag}^k$ are identified with the
tachyons $T_{k/R}, T_{-k/R}$ of momentum $\pm \frac{k}{R}$
respectively.

Since the matrix $Z$ commutes with its adjoint, the two can be
simultaneously diagonalised. The diagonalising matrices drop out of
the action leaving behind Vandermonde factors. It turns out that one
gets a single power of the Vandermonde for the eigenvalues 
$z_1,z_2,\ldots,z_N$ of $Z$, together with its complex
conjugate corresponding to $Z^\dagger$. Thus, for example, the
partition function at $t_k=\tbar_k=0$ is:
\beq
\label{znmmeigenvalues}
{\cal Z}_{NMM} = \int \prod_{i=1}^N d^2z_i\,\prod_{i<j}|z_i-z_j|^2 
~
e^{-\n \sum_{i=1}^N (z_i\bar{z}_i)^R  
 + \left(R\n-N +
\frac{R-1}{2}\right)\sum_{i=1}^N
\log z_i\bar{z}_i}
\eeq
with an obvious generalisation to include the tachyon perturbations.

At $t_k={\bar t}_k=0$, it can be shown (though not directly from the
action) that the NMM is invariant under the T-duality operation:
\beq
R\to\frac{1}{R},\qquad \mu\to\mu\,R
\eeq
This invariance is broken by the presence of momentum modes.  Indeed,
after T-duality, the tachyons $T_{\pm k/R}$ of the $c=1$ string turn
into winding modes of $\pm k$ units of winding, or equivalently (in
vertex-operator language) of left/right momentum
$(p_L,p_R)=\pm(kR,-kR)$. In what follows, these modes will be denoted
$\cT_{kR},\cT_{-kR}$.

In \cite{Alexandrov:2003qk} two distinct equivalences between the NMM
and the $c=1$ string were proposed. The first, referred to as ``Model
I'', requires us to take the large-$N$ limit of the NMM. The result in
this case was that:
\beq
\label{m1}
{\cal Z}_{c=1}(\mu,t,\tbar)=\lim\limits_{N \to \infty}
{\cal Z}^{(N)}_{NMM}(\nu,t,\tbar),
\eeq
after the analytic continuation $\nu=-i\mu$. 

However, another equivalence, ``Model II'', was proposed which did not
involve a large-$N$ limit. It was argued that the $c=1$ string theory
can be obtained from the NMM at \emph{finite} $N$, provided $\nu$ is
set to the special value $\frac{N}{R}$ (note that this corresponds to
an imaginary cosmological constant):
\beq
\label{m2}
{\cal Z}_{c=1}\left(\mu=i\sfrac{N}{R},t,\tbar\right)
={\cal Z}^{(N)}_{NMM}\left(\nu=\sfrac{N}{R},t,\tbar\right),
\eeq
In other words, the claim\footnote{The authors of
Ref.\cite{Alexandrov:2003qk} stated this a little differently: that
one obtains $c=1$ string amplitudes as a function of $\mu$ by
computing NMM correlators as a function of $N$ and $\mu$, and then
continuing $N$ to the imaginary value $-i\mu R$. This procedure is
less well-defined, as it requires us to make a discrete parameter
continuous.} is that an NMM calculation for a fixed integer value of
$N$ determines ${\cal Z}_{c=1}$ for a particular (imaginary) value of
$\mu$, namely
\beq
\mu = i\,\frac{N}{R}
\eeq
If we T-dualise the above
considerations so that $t,\tbar$ become couplings to winding tachyons,
this relation becomes
\beq
\mu = iN
\eeq
The above results seem to indicate that for finite $N$ we can only
generate the answer at a fixed $\mu$, in which case we would never
obtain the perturbative expansion in powers of $1/\mu^2$. However,
below we will compute winding correlators using the NMM, and will see
that it turns out much more powerful than expected. It actually does
reproduce the entire perturbative correlators, as functions of $\mu$
and $R$, even at finite values of $N$. Evidence for this fact, as well
as an explanation of it, will be provided in subsequent sections.

\section{Correlators from Matrix Quantum Mechanics}
\label{mqmcorr}

\subsection{Two-point functions}

We start by presenting formulae for the two-point function $\langle
T_{-n/R}T_{n/R}\rangle$ to all orders in $\frac{1}{\mu^2}$, from the
Matrix Quantum Mechanics (MQM) approach. We will derive these formulae,
valid at arbitary radius, starting from the infinite-radius formulae
presented in
\cite{Moore:1991sf}. We start by quoting the closed-form expression for
the infinite-radius two-point function $\langle T_{-q}T_q\rangle$, or
more precisely the first derivative of the two-point function with
respect to the cosmological constant, which is actually more
convenient for our purposes:
\beq
\label{ptmqtqinf}
\del_\m\langle T_{-q} T_q \rangle_\infty = 
\left(\Gamma(-q)\right)^2~{\rm Im}~ e^{i\pi
q/2}\left(\frac{\G\left(\half -i\m +q\right)}{\G\left(\half
-i\m\right)}-\frac{\G\left(\half -i\m\right)}{\G\left(\half
-i\m -q\right)}\right),
\eeq
where $q>0$. For clarity of presentation we will drop the leg-pole
factors $\left(\Gamma(-q)\right)^2$ in what follows, keeping in mind
that they can be restored whenever needed.

Now we obtain the corresponding amplitudes at a finite radius R, using
Eqs.(\ref{finiteR}) and (\ref{deltafactor}):
\beq
\label{ptmqtq}
~
\nn \langle T_{-q} T_q \rangle_R = R\,\frac{\frac{1}{2R}}{{\rm
sin}\left(\frac{1}{2R}\del_\m\right)}~{\rm Im}~ e^{i\pi
q/2}\left(\frac{\G\left(\half -i\m+q\right)}{\G\left(\half
-i\m\right)}-\frac{\G\left(\half -i\m\right)}{\G\left(\half
-i\m -q\right)}\right)
\eeq
where the first factor of $R$ comes from the replacement of the
$\delta$-function by a Kronecker $\delta$ as in \eref{deltafactor}.
The differential operator in front is real and acts only on functions
of $\m$, so it can be moved inside and we thus need to evaluate
\beq
\nn \frac{1}{2\,{\rm sin}\left(\frac{1}{2R}\del_\m\right)}
\left(\frac{\G\left(\half -i\m +q\right)}{\G\left(\half
-i\m\right)}-\frac{\G\left(\half -i\m\right)}{\G\left(\half
-i\m-q\right)}\right)
\eeq
This can be done very easily by expanding the operator as follows
\beq
\label{oper}
\nn \frac{1}{2\,{\rm sin}\left(\frac{1}{2R}\del_\m\right)} =
-i \sum_{j=0}^\infty e^{i\left(j +
\half\right)\frac{1}{R}\del_\m}
\eeq
Using this we get the required expression as
\beq
\label{expr2pt}
-i \sum_{j=0}^\infty \left(\frac{\G\left(\half -i\m + q+
\frac{j}{R} + \frac{1}{2R}\right)}{\G\left(\half -i\m + \frac{j}{R} +
\frac{1}{2R}\right)}-\frac{\G\left(\half -i\m + \frac{j}{R} +
\frac{1}{2R}\right)}{\G\left(\half -i\m -q + \frac{j}{R} +
\frac{1}{2R}\right)}\right)
\eeq
Next, we choose $q=n/R$. We see that the $j^{\rm th}$ term from the
first sum cancels the $(j+n)^{\rm th}$ term from the second sum. So
only the $j=0,1,\ldots,n-1$ terms from the second sum remain. Defining
$r=n-j$, the above expression becomes\footnote{Here and in what
follows, we drop the $R$ subscript in the correlators wherever it is
obvious that they are at finite $R$.}:
\beq
\langle T_{-n/R} T_{n/R}\rangle = {\rm Re}~ e^{i\pi
n/2R}\,\sum_{r=1}^n \frac{\G\left(\half -i\m + (r - \half)\frac{1}{R}
\right)}{\G\left(\half -i\m + (r - n - \half)\frac{1}{R}\right)}
\eeq

In order to obtain the expansion of this expression in powers of
$1/\mu^2$, we can rewrite it in terms of the special functions:
\beq
\cF^\pm(a,b;\mu) \equiv \frac{\Gamma(\half-i\mu +
a)}{\Gamma(\half-i\mu+b)} \pm
\frac{\Gamma(\half-i\mu -b)}{\Gamma(\half-i\mu-a)}
\label{functionf}
\eeq
defined in Eq.(B.2) of Ref.\cite{Moore:1991sf}. We have:
\bea
&&\langle T_{-n/R} T_{n/R}\rangle = {\rm Re}~ 
e^{i\pi n/2R} \sum_{r=1}^{n/2}\cF^+\Big((r-\shalf)\sfrac{1}{R},
(r-n-\shalf)\sfrac{1}{R};\mu\Big),\quad
~n~\hbox{even}\\
&&= {\rm Re}~ e^{i\pi n/2R} \Bigg(\sfrac12\cF^+
\left(\sfrac {n}{2R},-\sfrac {n}{2R};\mu\right) +
\sum_{r=1}^{(n-1)/2}\!\!\!
\cF^+\Big((r-\shalf)\sfrac1{R},(r-n-\shalf)\sfrac1{R};\mu\Big)
\Bigg),~n~\hbox{odd}\nn
\eea
Next we use the asymptotics for large $\mu$:
\beq
\label{fplusasymp}
\cF^+(a,b;\mu) = e^{-i\pi(a-b)/2}\,\mu^{a-b}\,f(a,b;\mu)
\eeq
where $f(a,b;\mu)$ is a power series in $\frac1{\mu^2}$ with real
coefficients and starting with a constant term:
\beq
f(a,b;\mu) = 2 - \sfrac1{12}
(a-b)(a-b-1)\left(3(a+b)^2-(a-b)-1\right)\frac1{\mu^2} + {\cal
O}\left(\frac1{\mu^4}\right) 
\eeq 

It follows that, for even $n$:
\bea
\label{wtmntn2}
\langle T_{-n/R} T_{n/R}\rangle &=& {\rm Re}~
\mu^{n/R}\sum_{r=1}^{n/2}f\Big((r-\shalf)\sfrac1{R},
(r-n-\shalf)\sfrac1{R};\mu\Big)\nn\\
&=& \mu^{n/R}
\sum_{r=1}^{n/2}f\Big((r-\shalf)\sfrac1{R},(r-n-\shalf)\sfrac1{R};\mu\Big)\nn\\
&=& \left|\sum_{r=1}^n \frac{\G(\half-i\mu 
+ (r-\shalf)\sfrac1{R})}
{\G(\half-i\mu + (r-n-\shalf)\sfrac1{R})}\right|
\eea
The first step above follows because the function $f$ is real. The
final equality is true for all $n$, and not just even values. This
then is the complete answer for the perturbative expansion of
two-point functions of momentum correlators at arbitrary radius.

Specialising to $n=1$, we find the following expression, which will
be useful later on:
\beq
\label{wcorr}
\langle T_{-1/R} T_{1/R}\rangle = \left|\frac{\G
\left(\half -i\m + \frac{1}{2R}
\right)}{\G\left(\half -i\m - \frac{1}{2R}\right)}\right|
\eeq
After a T-duality
\beq
R\to 1/R,\quad \mu\to\mu R
\eeq
we get the unit-winding two-point function
\beq
\label{unitwinding2pt}
\langle \cT_{-R} \cT_{R}\rangle = \left|\frac{\G
\left(\half -i\m R + \frac R2
\right)}{\G\left(\half -i\m R - \frac R2\right)}\right|
\eeq
This expression was recently derived by
Maldacena\cite{Maldacena:2005hi}. We should note that the above answer
has to be multiplied by the leg pole factor $\big(\G(-R)\big)^2$, which
we dropped after \eref{ptmqtqinf}.

\subsection{Four-point functions}
\label{4pfunc}

In this section we turn to the computation of higher point functions.
In particular, we extend the results for the four-point function from
MQM  to finite $R$ and then specialise to the
case of unit winding modes. In this case we will be able to find an
explicit all-orders result after summing an infinite series.

Upto leg pole factors (which can be unambiguously restored when
needed) the connected four-point function at infinite 
radius is\cite{Moore:1991sf}:
\bea
\label{ptmqtq2inf0}
~
\nn \del_\m\langle (T_{-q} T_q)^2 \rangle^{\rm conn}_\infty 
&=& {\rm Im}~ e^{i\pi
q}\left[{\cal F}^+(2q,0;\m)-{\cal F}^+(q,-q;\m) + \sum_{n=1}^\infty
\frac{(-1)^n}{n!}2\left(\frac{\G(-q+n)}{\G(-q)}\right)^2\right.\\
~
&& \times \left.\left(\frac{\G\left(2q - n + \half
-i\m\right)}{\G\left(\half -i\m\right)}-\frac{\G\left(q - n + \half
-i\m\right)}{\G\left(-q + \half -i\m\right)}\right)\right],
~
\eea
where $q>0$ and the function ${\cal F}^+$ is defined in
\eref{functionf}.

Substituting \eref{functionf} in \eref{ptmqtq2inf0} we have
\bea
\label{ptmqtq2inf}
~
&& \nn \del_\m\langle (T_{-q} T_q)^2 \rangle^{\rm conn}_\infty 
= {\rm Im}~ e^{i\pi
q}\left[\frac{\G\left(\half
-i\m+2q\right)}{\G\left(\half -i\m\right)}+\frac{\G\left(\half
-i\m\right)}{\G\left(\half -i\m-2q\right)} -2\frac{\G\left(\half
-i\m+q\right)}{\G\left(\half -i\m -q\right)} \right.\\
~
&& \nn \left. + 2\sum_{n=1}^\infty
\frac{(-1)^n}{n!}\left(\frac{\G(-q+n)}{\G(-q)}\right)^2 
\left(\frac{\G\left(\half
-i\m+2q-n\right)}{\G\left(\half -i\m\right)}-\frac{\G\left(\half
-i\m+q-n\right)}{\G\left(\half -i\m-q\right)}\right)\right] \\
~
\eea
The connected finite-$R$ amplitude is, therefore
\bea
\label{ptmqtq2}
~
\nn \langle (T_{-q} T_q)^2 \rangle^{\rm conn}_R 
&=& R\,\frac{\frac{1}{2R}\del_\m}{{\rm
sin}\left(\frac{1}{2R}\del_\m\right)}
\langle (T_{-q} T_q)^2 \rangle^{\rm conn}_\infty
~
\eea
We use the expansion \eref{oper} of the differential operator 
and set $q=1/R$ to get
\bea
\label{ptm1t12}
\nn \langle (T_{-1/R} T_{1/R})^2 \rangle^{\rm conn} 
&=& {\rm Re}~
e^{i\pi/R}
\left(-\frac{\G\left(\half - i\m +
\frac{3}{2R}\right)}{\G\left(\half - i\m - \frac{1}{2R}\right)}
+ \frac{\G\left( \half - i\m + 
\frac{1}{2R}\right)}{\G\left(\half - i\m - \frac{3}{2R}\right)} \right.\\
~
&& \left. -2\sum_{n=1}^\infty
\frac{(-1)^n}{n!}\left(\frac{\G(-\frac{1}{R}+n)}{\G(-\frac{1}{R})}\right)^2
\frac{\G\left(\half - i\m +
\frac{3}{2R} -n\right)}{\G\left(\half - i\m - \frac{1}{2R}\right)}
\right)
~
\eea
It is convenient to add and subtract a term corresponding to $n=0$ in
the summation. This extends the sum from $0$ to $\infty$, while the
subtracted term changes the sign of the first term above, after which
the first two terms combine into an $\cF^+$. Thus we get:
\bea
\label{ptm1t13}
~
\nn \langle (T_{-1/R} T_{1/R})^2 \rangle^{\rm conn} 
&=& {\rm Re}~
e^{i\pi/R}
\Bigg(\cF^+(\sfrac{3}{2R},-\sfrac{1}{2R};\mu)
\\
~
&& \left. -2\sum_{n=0}^\infty
\frac{(-1)^n}{n!}\left(\frac{\G(-\frac{1}{R}+n)}{\G(-\frac{1}{R})}\right)^2
\frac{\G\left(\half - i\m +
\frac{3}{2R}-n\right)}{\G\left( \half - i\m - \frac{1}{2R}\right)}
\right)
~
\eea
The sum is now easy to evaluate using the integral representations for
the three $\Gamma$-functions in the numerator that depend on $n$ (see
Appendix \ref{evalsum}). This finally leads to:
\bea
\label{ptm1t13-summed}
\nn
\langle (T_{-1/R} T_{1/R})^2 \rangle^{\rm conn} 
&=& {\rm Re}~ e^{i\pi/R}
\left(\cF^+(\sfrac{3}{2R},-\sfrac{1}{2R};\mu)
-2 \Big(\shalf \cF^+(\sfrac{1}{2R}, - 
\sfrac{1}{2R};\mu)\Big)^2
\right)\\
&=& \left|\cF^+(\sfrac{3}{2R},-\sfrac{1}{2R};\mu)
-2 \Big(\shalf \cF^+(\sfrac{1}{2R}, - 
\sfrac{1}{2R};\mu)\Big)^2
\right|
\eea
One can verify that the two terms above are, respectively, the full
(connected plus disconnected) correlator, and its disconnected
part.

\section{Correlators in the finite-$N$ Normal Matrix Model}
\label{nmmcorr}

Having obtained explicit expressions for all two-point and a
particular four-point function from the MQM, as a function of the
cosmological constant $\mu$ and radius $R$, we now attempt to recover
the same results from the NMM. This first of all provides a test of
the NMM and its effectiveness. But once we explore the systematics it
will become clear that we can compute much more. In fact, we will
obtain a complete combinatorial formula for the $2n$-point functions
of unit-momentum correlators. Via T-duality, this determines the
corresponding winding correlators. We expect this to be useful in
determining the full vortex condensate to all orders in perturbation
theory.

As mentioned before, in the process of studying the NMM we will
encounter a rather surprising result: for the purpose of computing
correlators, one can actually take $N$ to be a small finite value and
yet obtain the correct answer {\it as a function of $\mu$}. The finite
value of $N$ will be determined by the operators whose correlators we
are calculating. For this purpose it is convenient to classify tachyon
correlators into sectors labelled by an integer, the total positive
momentum $P$ flowing through that correlator, measured in units of
$1/R$. For example in $\langle T_{-k_1/R} T_{-k_2/R} T_{m_1/R}
T_{m_2/R}\rangle$, where $k_1,k_2,m_1,m_2$ are all positive, the total
positive momentum is $P=m_1+m_2 =k_1+k_2$. This number will determine
the minimum value of $N$ required in the NMM to compute these
correlators.  In what follows we will first consider all correlators
in the sectors $P=1$ and $P=2$. In the former case there is only a
single two-point function, while in the latter case we have two, three
and four-point functions. After presenting some examples we will
discuss why the theory works in this way.

\subsection{Two-point functions: examples}
\label{2ptfnex}

\nt{\it Example: $n=1$}
\medskip

We begin by computing the two point function of the unit momentum
operator. Since total momentum is conserved, this operator is paired
with the one of negative unit momentum. So we will calculate the two
point function $\langle T_{-1/R} T_{1/R}\rangle$ of unit momentum
operators.

We first calculate the partition function of NMM at $N=1$:
\beq
\label{zn1}
{\cal Z}_{NMM}^{N=1}(t=0)=\int dz d\bar{z}\,
e^{-\n (z\bar{z})^R + \left(R\n-1 +
\frac{R-1}{2}\right){\rm\,log}\,z\bar{z}}
\eeq
Setting $z=\sqrt{m}\,e^{i\te}$, $dz d\bar{z}\ra dm\,d\te$, we have:
\bea
\label{zn1full}
~
\nn {\cal Z}_{NMM}^{N=1}(t=0) &=& \int_0^\infty \int_0^{2\pi} dm\,d\te\,e^{-\n
m^R + \left(R\n-1 + \frac{R-1}{2}\right){\rm\,log}\,m} \\
~ 
\nn &=& 2\pi \int_0^\infty dm\,m^{\left(R\n-1 + \frac{R-1}{2}\right)} e^{-\n
m^R} \\ 
~ 
&=& \frac{2\pi}{R} \nu^{-\left(\n + \half - \frac{1}{2R}\right)}\G\left(\n +
\shalf - \sfrac{1}{2R}\right)
~  
\eea
As a function of $\nu$, this is not the correct partition function of
the $c=1$ string, but it reduces to the correct partition function if
in the above expression we set $\nu=\frac{1}{R}$ and compare this with
${\cal Z}_{c=1}(\frac{i}{R},t=0,\tbar=0)$. This fact is a direct
consequence of the claim in Ref.\cite{Alexandrov:2003qk}, see
\eref{m2}. It is also worth noting that the partition function at
$N=1$ is not invariant under T-duality.  In fact, T-duality in
the NMM partition function is recovered only in the limit
$N\to\infty$. This makes it clear that the correct partition function,
as a function of $\mu$ and $R$, can never be recovered at finite $N$.

For correlators, things are quite different, as we will now see. For
the two-point function, we find:
\bea
\label{dm1d1zn1}
~
\nn \del_{-1}\del_1 {\cal Z}_{NMM}^{N=1}(t=0) &=& \int_0^\infty \int_0^{2\pi}
dm\,d\te\,m\,e^{-\n m^R + \left(R\n-1 + \frac{R-1}{2}\right){\rm\,log}\,m} \\
~ 
&=& \frac{2\pi}{R} \nu^{-\left(\n + \half + \frac{1}{2R}\right)}\G\left(\n +
\shalf + \sfrac{1}{2R}\right)
~ 
\eea
{}From \eref{zn1full} and \eref{dm1d1zn1} we have:
\beq
\label{2point}
\del_{-1}\del_1 \ln{\cal Z}_{NMM}^{N=1}(t=0)
=\n^{-\frac{1}{R}}\frac{\G\left(\n
+ \half + \frac{1}{2R}\right)}{\G\left(\n + \half - \frac{1}{2R}\right)}
\eeq
Finally, we have to analytically continue $\n=-i\m$. The result is
complex, but can easily be seen to have the form of an overall phase
times a real power series in $1/\mu^2$. Dropping the phase is then
equivalent to taking the modulus of the above expression. This gives:
\beq
\label{ptm1t1}
\langle T_{-1/R} T_{1/R} \rangle_{NMM}^{N=1}=\mu^{-\frac{1}{R}}
\left|\frac{\G\left(\half - i\m +
\frac{1}{2R}\right)}
{\G\left(\half - i\m -
\frac{1}{2R}\right)}\right|
\eeq
which agrees with \eref{wcorr} upto the prefactor, $\mu^{-1/R}$,
which indicates that the ``tachyons'' of the NMM are normalised differently
from those of MQM. Indeed we will argue later that the relationship
is:
\bea
T_{n/R}|_{NMM} = \mu^{-n/2R}\,T_{n/R}|_{MQM}
\label{nmmnorm}
\eea

We have discovered the surprising result that the exact two-point
correlator of unit momentum tachyons is correctly calculated (as a
function of $\mu$ and $R$) using only the $1\times 1$ Normal Matrix
Model! According to \eref{m2}, we should have expected the result
to be correct only for $\mu = i/R$. We will see that a similar
feature holds for all two-point correlators, though the minimum
required value of $N$ depends on the correlator under
consideration. Later we will extend this observation to higher-point
correlators.
\bigskip

\nt{\it Example: $n=2$}
\medskip

We consider another example, the correlator $\langle T_{-2/R}
T_{2/R}\rangle$. In this case, according to the prediction in
\eref{m2}, we can perform a calculation at $N=1$ and the result so
obtained will be valid at the special value of the cosmological
constant $\nu = 1$. However, we now face a puzzle. In the NMM at
$N=1$, one cannot distinguish the four correlators:
\beq
\langle T_{-2/R}\, T_{2/R}\rangle, \quad
\langle T_{-2/R}\, T_{1/R}\, T_{1/R}\rangle, \quad
\langle T_{-1/R}\, T_{-1/R}\, T_{2/R}\rangle, \quad
\langle T_{-1/R}\, T_{-1/R}\,T_{1/R}\, T_{1/R}\rangle 
\label{4corr}
\eeq
because all of these are represented by the same NMM
correlator $\langle z^2 \zbar^2\rangle$. Therefore, assuming
\eref{m2} continues to hold, either it has to be the case that all
four correlators become the same at $\nu =\frac{1}{R}$, or else at
best we can only hope to obtain some linear combination of them.

The calculation is straightforward and upon continuing to $\nu=-i\mu$ 
and taking the modulus, we find:
\beq
\langle T_{-2/R}T_{2/R}\rangle_{NMM}^{N=1}
=\mu^{-2/R}\left|\frac{\G\left(\half -i\mu + \frac{3}{2R}\right)}
{\G\left(\half -i\mu - \frac{1}{2R}\right)}\right|
\label{nmmtm2t2neq1}
\eeq
This can be compared with the known result from
\eref{wtmntn2}. Specialising to the present case, and changing to the
NMM normalisation via \eref{nmmnorm} gives us:
\beq
\langle T_{-2/R} T_{2/R}\rangle = 
\left|\frac{\G(\half-i\mu +\frac{1}{2R})}{\G(\half-i\mu -\frac{3}{2R})} +
\frac{\G(\half-i\mu +\frac{3}{2R})}{\G(\half-i\mu
-\frac{1}{2R})}\right|
\label{mqmtm2t2}
\eeq
Comparing Eqs.(\ref{nmmtm2t2neq1}),(\ref{mqmtm2t2}), we see that the
NMM result for this correlator at $N=1$ is not correct. This is not a
surprise. But now we see that it is incorrect even at the special
value $\mu = i/R$, which appears to contradict \eref{m2}. As we will
see, this is due to the fact that the same NMM correlator can describe
different tachyon correlation functions for low $N$. Indeed, one can
check that the answer we have obtained at $N=1$ in
\eref{nmmtm2t2neq1} is actually a linear combination of the
correlators in \eref{4corr} as calculated from matrix quantum
mechanics.

Let us continue by evaluating the NMM correlator at $N=2$. In this
case the operator we are dealing with is
$T_{2/R}\sim \tr Z^2$ which is linearly independent of
$(T_{1/R})^2\sim (\tr Z)^2$ once $Z$ is a $2\times 2$ matrix, so there
is no longer a risk of mixing for the operators in \eref{4corr}.
The computation is given in an Appendix, and leads to the answer
\eref{ptm2t2}, which after changing to the NMM normalisation is:
\beq
\langle T_{-2/R}T_{2/R}\rangle_{NMM}^{N=2}=
\left|\frac{\G\left(\half - i\m  +
\frac{3}{2R}\right)}{\G\left(\half - i\m - \frac{1}{2R}\right)} +
\frac{\G\left(\half - i\m  + \frac{1}{2R}\right)}{\G\left(\half - i\m 
- \frac{3}{2R}\right)}\right|
\eeq
Following \eref{m2} we would expect that this should give the correct
answer for $\mu=2i/R$. But now there is a surprise, since in fact it
agrees perfectly with the MQM result \eref{mqmtm2t2} for {\it all}
values of $\mu$. Thus for the purposes of calculating $\langle
T_{-2/R} T_{2/R}\rangle$ in $c=1$ string theory, to all orders in the
string coupling, a $2\times 2$ matrix model is sufficient.

To summarise, we have found evidence that an NMM calculation of
tachyon correlators at finite $N$ (where the minimum required value of
$N$ depends on the correlator in question) gives the correct tachyon
correlators for the $c=1$ string, to all orders in perturbation
theory. Below we will collect more evidence for this property, which
appears to go far beyond the result of Ref.\cite{Alexandrov:2003qk} as
stated in \eref{m2} above.

\subsection{Two-point functions: general case}
\label{2ptfn}

Let us now consider the general case $\langle T_{-n/R} T_{n/R}\rangle$. 
and try to derive this result from the NMM. We will find
that for this correlator, the NMM with $N=n$ is
sufficient to give the correct result. Indeed, when we compute in the
$N\times N$ NMM starting at $N=1$ and increasing $N$ in integer steps,
we obtain the right $c=1$ string correlator (as a function of $\m$) 
as long as $N\ge n$, though not for $N<n$. Thus the NMM calculation
``stabilises'' at a certain minimum value of $N$.

Since we will be computing normalised correlators, we start by
computing the (unperturbed) partition function at a general value of
$N$. This is given by
\bea
\label{znnfull1}
~
&& {\cal Z}_{NMM}^{N}(t=0) = \int_0^\infty \prod_{r=1}^{N} dm_r
\int_0^{2\pi} \prod_{r=1}^{N} d\te_r \\
~
\nn && \times \prod_{j<k}^N \left(m_j + m_k - \sqrt{m_j m_k}(e^{i\te_{jk}} +
e^{-i\te_{jk}})\right) e^{-\n\left(\sum_{r=1}^{N}m_r^R\right) + \left(R\n-N +
\frac{R-1}{2}\right)\left(\sum_{r=1}^{N} {\rm log}\,m_r\right)}
~
\eea
The next step is to perform the integration over the $\te$'s. In
general this will be quite tedious, because one has to pick out terms
which are independent of $\te$ by expanding out the Vandermonde
factor. However, we notice that since the above expression is
invariant under permutations of $m$'s, we can determine all terms
surviving the $\te$ integrals if we know just one of them, by
permuting the $m$'s among themselves.

The first such term is just the product of the first term from each of the
Vandermonde factors, which is $m_1^{N-1}m_2^{N-2}\cdots m_{N-1}$. Thus we have,
after evaluating the $\te$ integrals
\bea
\label{znnfull2}
~
&& {\cal Z}_{NMM}^{N}(t=0) = (2\pi)^N N! \int_0^\infty \prod_{r=1}^{N}
dm_r \prod_{j=1}^{N-1} m_j^{N-j} \\
~
\nn && \hspace{4.5cm}\times~ 
e^{-\n\left(\sum_{r=1}^{N}m_r^R\right) + \left(R\n-N +
\frac{R-1}{2}\right)\left(\sum_{r=1}^{N} {\rm log}\,m_r\right)} \\
~
\nn && \hspace{2.5cm}= (2\pi)^N N! 
\prod_{r=1}^N \n^{-\left(\n + \half - \left(r -
\half\right)\frac{1}{R}\right)} \G\Big(\n + \shalf - \left(r -
\shalf\right)\sfrac{1}{R}\Big)
\eea
>From now on we will restrict to the case $N=n$.

The next step is to compute the two point function and then normalise
by the above partition function. We have:
\bea
~
&& \nn \del_{-n}\del_n {\cal Z}_{NMM}^{N=n}(t=0) =  
\int \prod_{r=1}^{n} d^2z_r
\prod_{j<k}^{n} |z_j-z_k|^2 
\left(\sum_{l=1}^n z_l^n\right)\left(\sum_{l=1}^n
\bar{z}_l^n\right) \\
~
&& \nn\hspace{3.6 cm} \times e^{-\n 
\left(\sum_{r=1}^n(z_r\bar{z}_r)^R\right) + \left(R\n-n +
\frac{R-1}{2}\right)\left(\sum_{r=1}^n{\rm\,log}\,z_r\bar{z}_r\right)} \\
~
&& \nn = \int_0^\infty 
\prod_{r=1}^{n}dm_r \int_0^{2\pi} \prod_{r=1}^{n}d\te_r
\prod_{j<k}^{n}\left(m_j + m_k - \sqrt{m_j m_k}(e^{i\te_{jk}} +
e^{-i\te_{jk}})\right) \\
~
&& \nn  \times 
\left(\sum_{r=1}^n \left(\sqrt m_r\right)^n e^{in\te_r}\right)
\left(\sum_{r=1}^n \left(\sqrt m_r\right)^n e^{-in\te_r}\right) \\
~
&& \times e^{-\n \left(\sum_{r=1}^n m_r^R\right) + \left(R\n-n +
\frac{R-1}{2}\right)\left(\sum_{r=1}^n{\rm\,log}\,m_r\right)}
~
\eea
In this case also we can avoid tedious calculation by applying the
permutation trick. The contribution to the first term from the
Vandermonde is same as before, and the contribution from $\tr Z^n\,
\tr Z^{\dag n}$ is $\sum_{r=1}^n m_r^n$. The net contribution is then
$\left(\sum_{r=1}^n m_r^n\right)m_1^{n-1}m_2^{n-2}\cdots
m_{n-1}$. Proceeding as before we have after the $\te$ integrals
\bea
\label{dmmndmnznn}
~
\nn && \del_{-n}\del_n {\cal Z}_{NMM}^{N=n}(t=0) = (2\pi)^n n! \int_0^\infty
\prod_{r=1}^{n} dm_r \left(\sum_{r=1}^n m_r^n\right) \prod_{j=1}^{n-1}
m_j^{n-j} \\
~
\nn && \hspace{3.6 cm}\times e^{-\n\left(\sum_{r=1}^{n}m_r^R\right) +
\left(R\n-n + \frac{R-1}{2}\right)\left(\sum_{r=1}^{n} {\rm log}\,m_r\right)}
\\
~
\nn &&= (2\pi)^n n! \sum_{j=1}^n\left[\n^{-\left(\n + \half - \left(j - n -
\half\right)\frac{1}{R}\right)}\G\left(\n + \half - \left(j - n -
\half\right)\frac{1}{R}\right)\right. \\
~
&& \times \prod_{_{r\ne j}^{r=1}}^n \n^{-\left(\n + \half - \left(r -
\half\right)\frac{1}{R}\right)} \left.\G\left(\n + \half - \left(r -
\half\right)\frac{1}{R}\right)\right]
~
\eea
{}From \eref{znnfull2} and \eref{dmmndmnznn} we find (after changing
variables $j\ra n+1-r$):
\beq
\label{ptmntntmp}
\langle T_{-n/R} T_{n/R}\rangle_{NMM}^{N=n}=\n^{-n/R}\sum_{r=1}^n
\frac{\G\left(\half -i\m + (r - \half)\frac{1}{R} \right)}{\G\left(\half -i\m
+ (r - n - \half)\frac{1}{R}\right)}
\eeq
As before, we analytically continue $\n=-i\m$ and take the modulus to get:
\beq
\label{ptmntn}
\langle T_{-n/R} T_{n/R}\rangle_{NMM}^{N=n}=\mu^{-n/R}\left| \sum_{r=1}^n
\frac{\G\left(\half -i\m + (r - \half)\frac{1}{R} \right)}{\G\left(\half -i\m
+ (r - n - \half)\frac{1}{R}\right)}\right|
\eeq
After changing normalisation via \eref{nmmnorm}, we see that this
agrees perfectly with \eref{wtmntn2}.

The above calculation was performed with matrices of rank $N=n$.  It
can easily be repeated for the other cases. When $N$ is smaller than
$n$, we find that the answer, as a function of $\m$, is not equal to
the correct two-point function, and does not become the correct one
even after choosing $\mu = in/R$. As before, this is due to
``contamination'' by correlators of higher point functions carrying
the same total momentum, because for $N<n$ the corresponding
correlators in the NMM are not all linearly independent.  For $N> n$,
instead, we actually get the {\it same} final answer as for $N=n$. The
calculational procedure we described above seems to suggest that extra
terms arise for $N>n$, but actually they are cancelled by
contributions from the $\te$ dependent terms in the Vandermonde
factor. Thus when we take the ratio of $\del_{-n}\del_n {\cal Z}$ and
${\cal Z}$ we end up with the RHS of
\eref{ptmntntmp}. Therefore as long as we take $N\ge n$, we get the
right answer (independent of $N$) for every $N$. This is what we
referred to as ``stabilisation'' above.

\subsection{Four-point functions}

Now we would like to compute the four-point function in the Normal
Matrix Model. For $N=1$, the calculation has already been performed,
since as we noted above, it is the same as the corresponding
calculation for the two-point function in \eref{nmmtm2t2neq1} (more
precisely the disconnected four-point function is the same as this
two-point function). As we explained there, the result so obtained is
a linear combination of the correct two, three and four-point
functions of the $c=1$ string, and to distinguish them we need to go
to a higher value of $N$. Accordingly we have computed the above
four-point function using the $N=2$ NMM. The derivation can be found
in Appendix \ref{fourpoint}, and the result is:
\beq
\label{nmmptm1t13}
\langle (T_{-1/R} T_{1/R})^2 \rangle_{NMM}^{N=2} = \mu^{-2/R}
\left|\cF^+(\sfrac{3}{2R},-\sfrac{1}{2R};\mu)
-\shalf \Big(\cF^+(\sfrac{1}{2R}, - \sfrac{1}{2R};\mu)\Big)^2
\right|
\eeq
Changing from MQM to NMM normalisation using \eref{nmmnorm}, and
inserting the usual $1/R$ factor, we see that
\eref{nmmptm1t13} above is identical to \eref{ptm1t13-summed}.

For completeness, let us briefly consider the two three-point
functions 
\bea
\langle \cT_{-2R} \cT_{R}\cT_R
\rangle_{NMM}^{N=2},\qquad
\langle \cT_{2R} \cT_{-R}\cT_{-R}
\rangle_{NMM}^{N=2}
\eea
The two are actually equal to each other because of the symmetry $X\to
-X$, where $X$ is the Euclidean time direction.  We have calculated
these correlators both from MQM and NMM (at $N=2$) and the
agreement is exactly as for the cases considered above.

\subsection{Why it works}

As we reviewed in Section \ref{mqmnmmreview}, the Normal Matrix Model
determines every momentum correlator by differentiation with respect
to the momentum couplings $t,\tbar$. However, the correlators so
obtained should only be correct in the limit $N\to\infty$ (``Model
I'', \eref{m1}) or the special values $N=\nu R$ (``Model II'',
\eref{m2}). Now in the previous subsections we have shown in several
examples (including the infinite set of two-point functions) that,
given the total momentum $P$ flowing in the correlator, the NMM with
matrices of any rank $N\ge P$ suffices to compute the correlator {\it
completely} as a function of $\mu$ and $R$. In view of this, the NMM
appears to go beyond its expected range of validity. Here we will give
an explanation as to how this comes about.

The basic observation is that the phenomenon we are observing is not
to be viewed as an application of Model II, but rather of Model
I. Indeed, using Model II and a definite value of $N$, it is clear from
\eref{m2} that the answers obtained are correct only for a definite
value of $\nu$, namely $\nu=N/R$. This relation between $N$ and $\nu$
defines a line in $(N,\nu)$ space, and the points on this line where
$N$ takes integer values are the ones where the procedure
works. However, it is clear that in this way one can never recover the
full $\nu$ dependence at a fixed $N$.

In contrast, in Model I one is supposed to compute correlators at an
arbitrarily large value of $N$ and in the limit $N\to\infty$, the
correct answers are obtained as a function of $\nu$. What
we will now show is that, after computing a given correlator of total
momentum $P$ in this way, and then dividing by the partition function,
infinitely many terms cancel out exactly in the ratio. The remaining
terms, which actually contribute to the correlator of interest, are
the same as one would compute for a finite value of $N$, namely $N=P$.

The argument goes as follows. From the derivation we have given in the
previous subsections and the appendices, any correlator is generated
(after $\theta_i$ integrations) by inserting an expression of the form
$\prod_{i=1}^N m_i^{\al_i}$ into the $m_i$ integrals, where
$\{\al_i\}$ correspond to ordered partitions of $P$. Therefore we
should first of all choose $N$ large enough so that all such
partitions can be realised and are distinguishable. This is possible
for $N\ge P$. For $N<P$ we will miss some partitions, and thus the
answer cannot be correct. But the case $N>P$ realises the same
partitions as the case $N=P$ and thus gives the same answer. This
causes what we earlier called ``stabilisation'', which amounts to
saying that the result for $N=P$ is identical to the result for any
$N>P$, and therefore for $N=\infty$. Invoking the converse of
stabilisation, we can therefore start with the model defined at
$N=\infty$ and ``bring back'' the value of $N$ to any finite value
$N\ge P$ without changing the result. This explains why a finite-$N$
matrix model is sufficient to compute any momentum correlator.

\subsection{Combinatorial result for $2n$-point functions}
\label{combresult}

We have shown that the NMM is an effective tool by re-computing known
correlators. Now that we understand how and why it works, we apply it
to compute a new result: the full (connected plus disconnected)
$2n$-point function $\langle (T_{-1/R}T_{1/R})^n\rangle$ for every $n$
and to all orders in perturbation theory. The result, derived in
Appendix \ref{2npointfunctions}, is the following:
\beq
\label{2npt}
\langle(T_{-1/R}T_{1/R})^n\rangle = \left|\sum_{\{k_i\}}
C(\{k_i\})^2\prod_{i=1}^n \frac{\G\left(\shalf-i\m
+(k_i-n+\shalf)\sfrac{1}{R}\right)} {\G\left(\shalf-i\m
-(i-\shalf)\sfrac{1}{R}\right)}\right|
\eeq
with $C(\{k_i\})$ defined as:
\beq
C(\{k_i\}) = 
\sum_{\cP}(-1)^\cP 
\prod_{i=1}^n \ncr{n-\sum_{j=1}^{i-1}(k_j-\cP_j)}{k_i-\cP_i}
\eeq
Here, $\{k_i\}$ are strictly ordered partitions of $n(n+1)/2$, namely:
\beq
k_1>k_2>\cdots>k_n,\qquad \sum_{i=1}^n k_i = \frac{n(n+1)}{2}
\eeq
and $\cP$ denote permutations of the $n$ numbers $n-1,n-2,\cdots,0$.

Let us examine this result more closely. In principle, for every $n$
the answer is a sum of terms, each one being the ratio of $n$
$\G$-functions divided by $n$ $\G$-functions. However in practice,
some of the numerator and denominator terms can cancel out. We can see
this more explicitly if we list the first few special cases, of which
the first two have already been noted above:
\bea
\label{2nptex}
\nn\langle T_{-1/R}T_{1/R}\rangle &=& \left|\frac{\G\left(\half - i\m +
\frac{1}{2R}\right)} {\G\left(\half - i\m -
\frac{1}{2R}\right)}\right|\\
\langle(T_{-1/R}T_{1/R})^2\rangle &=&\left|\frac{\G\left(\half - i\m  +
\frac{3}{2R}\right)}{\G\left(\half - i\m - \frac{1}{2R}\right)} +
\frac{\G\left(\half - i\m  + \frac{1}{2R}\right)}{\G\left(\half - i\m 
- \frac{3}{2R}\right)}\right|\\
\nn\langle(T_{-1/R}T_{1/R})^3\rangle &=& \left|\frac{\G(\half-i\m +
\frac{5}{2R})}{\G(\half-i\m - \frac{1}{2R})} +
4 \frac{\G(\half-i\m +
\frac{3}{2R})}{\G(\half-i\m - \frac{3}{2R})}+
\frac{\G(\half-i\m +
\frac{1}{2R})}{\G(\half-i\m - \frac{5}{2R})}\right| 
\eea
The pattern emerging so far is misleadingly simple, as we see with the
next example, the 8-point function:
\bea
\nn\langle(T_{-1/R}T_{1/R})^4\rangle &=& \Bigg|\frac{\G(\half-i\m +
\frac{7}{2R})}{\G(\half-i\m - \frac{1}{2R})} +
9\, \frac{\G(\half-i\m + \frac{5}{2R})}{\G(\half-i\m - \frac{3}{2R})} +
9\, \frac{\G(\half-i\m + \frac{3}{2R})}{\G(\half-i\m -
\frac{5}{2R})}\\
&+& \frac{\G(\half-i\m + \frac{1}{2R})}{\G(\half-i\m - \frac{7}{2R})} +
4\, \frac{\G(\half-i\m + \frac{1}{2R})}{\G(\half-i\m - \frac{1}{2R})}
\frac{\G(\half-i\m + \frac{3}{2R})}{\G(\half-i\m - \frac{3}{2R})}\Bigg|
\eea
We see that as the number of operators in the correlator grows, one
gets products of more and more $\G$-functions in the numerator and
denominator. In this example we also see clearly that the coefficients
are perfect squares. 

Ideally one would like to know the connected part of the $2n$-point
function. In principle this can of course be obtained by repeated
application of \eref{2npt}, but one would like a more explicit and
useful expression. However, for the most likely application, to the
vortex condensate, we will not really need to make the distinction
between connected and disconnected correlators. The vortex condensate
corresponds to the partition function of a perturbed theory, and to
find the connected component of that it suffices to take a
logarithm. We will discuss this issue further in the following
section.

\section{Applications}
\label{applications}

\subsection{T-duality at $c=1$}

In this subsection we discuss how our results can be applied to check
T-duality of the $c=1$ matrix model. As we have seen, in the Euclidean
(finite-temperature) MQM, the momentum and winding modes with respect
to the time direction are independently defined. The former arise from
macroscopic loops defined in terms of fermion bilinears, while the
latter are Wilson-Polyakov loops in the thermal direction, which
project the theory onto nonsinglet sectors. From the continuum
description we expect that there should be T-duality between these two
sets of observables. Indeed, in Ref.\cite{Kazakov:2000pm} it has been
formally argued that, like the momentum-perturbed matrix model, the
winding-perturbed MQM also corresponds to the $\tau$-function of a
Toda hierarchy. To understand T-duality
better, one would like to compare explicit correlation functions
computed from the momentum and winding sides.

An attempt to directly check T-duality was made by Maldacena in
\cite{Maldacena:2005hi}, where the following two quantities were
compared: (i) the two-point function of unit-momentum tachyons, after
T-duality, and (ii) the partition function of MQM in the adjoint
sector. From \eref{unitwinding2pt} we see that (i) is equal to:
\beq
\langle \cT_{-R} \cT_{R}\rangle = \left|\frac{\G
\left(\half -i\m R + \frac R2
\right)}{\G\left(\half -i\m R - \frac R2\right)}\right|
\eeq
However, at this point we recall that leg-pole factors of $\G(-|q|)$
were dropped after \eref{ptmqtqinf}. Restoring them and taking the 
large-$\mu$ asymptotics of this correlator, we find\footnote{The
factor $R^R$ was not written in Ref.\cite{Maldacena:2005hi}.}:
\beq
\label{vortexrenorm}
\langle \cT_{-R} \cT_{R}\rangle = \big(\G(-R)\big)^2(\mu R)^R
\left(1+\frac{1}{24}\Big(R-\frac{1}{R}\Big)\mu^{-2}+
\cO\left(\mu^{-4}\right)\right)
\eeq
On the other hand, (ii) is obtained by solving MQM in the adjoint
sector. In the large $N$ limit, Maldacena obtained the leading (tree
level) contribution to the partition function in this sector as:
\beq
\label{adjpart}
\frac{Z_{\rm adj}}{Z_{\rm sing}} = \langle \cW_{\rm adj}\rangle=
\frac{1}{4\sin^2 \pi R}\,\mu^R = \frac{1}{4\pi^2}\,
\big(\G(R+1)\G(-R)\big)^2\mu^R
\eeq
The power of $\mu$ agrees with that in the leading term of
\eref{vortexrenorm}. The remaining discrepancy can be assigned to the
normalisation of the fundamental Wilson-Polyakov loop (or equivalently
to the normalisation of the original momentum modes), and we see that
Eqs.(\ref{adjpart}) and (\ref{vortexrenorm}) agree to leading order if
we change the normalisation of this loop variable to:
\beq
\label{wilsonrenorm}
\cW_{N} \to \frac{1}{2\pi}\,\frac{R^{\frac{R}{2}}}{\G(R+1)}
\cW_N
\eeq
This is a relatively simple change of normalisation\footnote{Notice
that the normalisation factor becomes trivial at the special radius
$R=1$.}, and appears to specify the basis in which T-duality holds in
MQM.

It is not entirely surprising that one needs to change normalisation
of the matrix model observables in order to implement
T-duality. Indeed, this duality is most manifest in the worldsheet or
Liouville approach, in which the momentum and winding vertex operators
come with a natural normalisation and are related to each other by the
simple change $(X_L,X_R)\to (X_L,-X_R)$. On the matrix model side,
momentum operators in the MQM are related to the corresponding
Liouville operators by a change of normalisation,
\eref{mqmnorm}. So one should expect that winding
operators in MQM are also related to Liouville winding operators by a
change of normalisation.

This is not to say we understand the nature of these normalisation
factors in general. In fact, as stressed in
Ref.\cite{Maldacena:2005hi}, we need more examples in order to check
the consistency of this picture. As an example, if one could compute
the genus-1 correction to the adjoint sector partition function, this
could be compared with the genus-1 term in \eref{vortexrenorm}. Similarly,
if one could compute the leading term for those higher representations
that correspond to $2n$-point functions of the winding tachyon, then
one could match this with the asymptotics of the latter, which can be
read off from our results in Section
\ref{combresult}.

There will also be representations corresponding to the
correlators of multiply wound tachyons $\cT_{nR}$. These correlators
can be found by T-dualising the relevant momentum correlators, for
example the two-point functions are found by T-dualising
\eref{wtmntn2}, leading to:
\bea
\label{nwindingcorr}
\langle\cT_{-nR}\cT_{nR}\rangle&=&
\big(\G(-nR)\big)^2 \left|\sum_{r=1}^n \frac{\G(\half-i\mu R 
+ (r-\shalf)R)}{\G(\half-i\mu R + (r-n-\shalf)R)}
\right|\\[2mm]
&&\hspace{-2.2cm}= n(\mu R)^{nR}\big(\G(-nR)\big)^2
\left(1- \frac{nR(nR-1)\big( (n^2-1)R^2 -nR
-1\big)}{24R^2}\mu^{-2} + \cO\left(\mu^{-4}\right)\right)
\nn
\eea
In the matrix model, this should correspond to the fundamental
Wilson-Polyakov loop with a contour that winds $n$ times over the time
direction. In principle we are allowed an independent choice of
normalisation for each winding number. In fact the momentum and
winding modes have corresponding freedoms in normalisation, and the
only thing relevant for T-duality is the relative normalisation
between them. So when we consider the nonsinglet sector related to
multiply wound loops, and the corresponding tachyons of $n$ units of
momentum, the leading-order comparison will be used to fix the
normalisation and the loop corrections will constitute a genuine check
of T-duality.

To summarise, we have not been able to address the problem of
T-duality but only set up one side of it. Namely, we have exhibited
the all-orders finite-radius correlators computed from the momentum
side, after performing a T-duality transformation. This constitutes a
prediction to be checked once it is properly understood how to perform
nonsinglet computations for different representations and to higher
orders in string perturbation theory.

There is one more intriguing point that we would like to mention.
The correlators we have computed take very special values at the
selfdual radius $R=1$, the point of enhanced $SU(2)$ symmetry. In
particular, all loop corrections to the two-point function of unit
momentum tachyons vanish, as can be seen from
\eref{unitwinding2pt}. Thus the tree level answer is
exact\footnote{This was already known long ago, for example as the
puncture equation in the Kontsevich-Penner
model\cite{Imbimbo:1995yv}.}. By T-duality the same property should
hold for the two-point function of unit winding modes. It is plausible
that one could extract this simple property just from the structure of
the nonsinglet Hamiltonian -- in this case it is the adjoint
Hamiltonian that was studied in Ref.\cite{Maldacena:2005hi},
specialised to $R=1$. Similarly, at $R=1$ the other two-point
functions have perturbation series that terminate at a finite number
of loops, as one can easily check from
\eref{nwindingcorr}. So, for consistency this
must also be a property of the antisymmetric-antisymmetric
representations referred to above. It may be simpler to derive this
kind of general result in the nonsinglet sector than to actually
compute coefficients with precision.

\subsection{Vortex condensate and black holes}

It is believed that the Euclidean 2D black hole background, defined in
the continuum by an $SL(2,R/U(1)$ CFT, is equivalent to the $c=1$
matrix model perturbed by fundamental Wilson-Polyakov loops:
\beq
S_{MQM}\to S_{MQM}+\lambda\,\cW_{N} + {\bar\lambda}\,\cW_{\bar N}
\eeq
The basis for this belief is the FZZ conjecture\cite{FZZconjecture},
which relates the black hole background to Sine-Liouville
theory\footnote{This conjecture has been proved by Hori and
Kapustin\cite{Hori:2001ax} in the ${\cal N}=2$ supersymmetric case. As
Maldacena has argued\cite{Maldacena:2005hi}, suitably orbifolding both
sides of their argument leads to a proof for the bosonic case.}. Via
the equivalence in \eref{wilsonwinding}, the latter is the same as the
perturbed background above.

To be precise, the FZZ conjecture is not really an either/or statement
wherein one uses either the black hole background or the
Sine-Liouville perturbation. It has increasingly become clear that the
backgrounds that one might call ``black hole'' or ``Sine-Liouville''
are the same, and both perturbations are turned on
simultaneously. Depending on the value of the worldsheet coupling, one
or the other of these perturbations is more dominant, but for example
the exact correlation functions have poles corresponding to both
perturbations\footnote{See for example Ref.\cite{Kutasov:2005rr}. We
are grateful to Ari Pakman for explaining this to us.}. In the
present work we will not focus on these details, but will be content
to treat the black hole story as a motivation to understand the vortex
condensate:
\beq
\langle e^{\lambda\,\cW_{N} + {\bar\lambda}\,\cW_{\bar
N}}\rangle\Big|_{MQM} 
\eeq
One way to compute this condensate would be to sum over an infinite
set of nonsinglet sectors in the MQM with some definite
weights. However, as we have seen, the technology to do this seems
rather limited at present. An alternative is to assume T-duality to
compute the correlator:
\beq
\langle e^{\lambda\,\cT_R + {\bar\lambda}\,\cT_{-R}}\rangle
=\sum_{n=0}^\infty \sum_{m=0}^\infty \frac{\lambda^n}{n!}
\frac{{\bar\lambda}^m}{m!}
\langle (\cT_{R})^n(\cT_{-R})^m\rangle =
\sum_{n=0}^\infty \frac{|\lambda|^{2n}}{(n!)^2}
\langle (\cT_{R}\cT_{-R})^n\rangle
\eeq
where the last equality follows from conservation of winding number.

Now from the computation in Appendix \ref{2npointfunctions}, we have
the following result after T-duality:
\beq
\langle(\cT_{-R}\cT_{R})^n\rangle = \left|\sum_{\{k_i\}}
C(\{k_i\})^2\prod_{i=1}^n \frac{\G\left(\shalf-i\m R
-(i+k_i-\shalf)R\right)} {\G\left(\shalf-i\m R
-(i-\shalf)R\right)}\right|
\eeq
where $\{k_i\}$ are strictly ordered partitions of $n(n+1)/2$, and
$C(\{k_i\})$ are the combinatorial coefficients given in \eref{ck}.

The above correlators contain both connected and disconnected
contributions. We can now pass to the generating function: 
\beq
\langle e^{\lambda\,\cT_R + {\bar\lambda}\,\cT_{-R}}\rangle
=\sum_{n=0}^\infty \frac{|\lambda|^{2n}}{(n!)^2}
\left|\sum_{\{k_i\}}
C(\{k_i\})^2\prod_{i=1}^n \frac{\G\left(\shalf-i\m R
-(i+k_i-\shalf)R\right)} {\G\left(\shalf-i\m R
-(i-\shalf)R\right)}\right|
\eeq
This is the partition function in the presence of a vortex condensate,
and its logarithm is the free energy of the perturbed theory. So one
does not need at any point to compute individual connected correlators.

The above expression is completely explicit and does not require
integrating any equation or developing a recursion relation. We expect
it will be useful to to extract physical quantities of interest
related to the Euclidean 2d black hole. This is beyond the scope of
the present work, however, and we hope to return to a more detailed
analysis of this formula in the future.

Again it is worth pointing out that at the selfdual radius $R=1$ the
vortex condensate is known exactly, though deriving it from the above
expression would not be the easiest way. The puncture equation of
Ref.\cite{Imbimbo:1995yv} simply tells us that:
\beq
\langle e^{\lambda\,\cT_R + 
{\bar\lambda}\,\cT_{-R}}\rangle|_{R=1} = |e^{-i\mu\lambda{\bar\lambda}}|
\eeq
and one can check easily that this agrees with the cases in
\eref{2nptex} specialised to $R=1$.

The significance for the Euclidean 2d black hole of this simple result
has not, to our knowledge, been explored. While it is true that the
black hole CFT corresponds to a radius $R=\sfrac{3}{2}$, it is
believed\cite{Kazakov:2000pm} to have a marginal deformation to other
radii at least in the range $1<R<2$. So the physical consequences of
the simple formula above at $R=1$ would be worth understanding better.

\section{Conclusions}
\label{conclusions}

In this work we have examined the familiar $c=1$ bosonic noncritical
string theory, or rather its Euclidean (finite temperature) version,
from the perspective of correlation functions. Both old and new
techniques were used to develop simple, elegant and explicit formulae
as functions of two variables: the cosmological constant $\mu$ and the
compactification radius or inverse temperature $R$. The key results
are summarised in
Eqs.(\ref{wtmntn2}),(\ref{ptm1t13-summed}),(\ref{2npt}). In addition
we have shown that the Normal Matrix Model is a powerful computational
tool.

An obvious extension of this work would be to the case of noncritical
type 0 strings\cite{Takayanagi:2003sm},\cite{Douglas:2003up}. In
Ref.\cite{Maldacena:2005he}, explicit expressions are obtained for the
partition functions of type 0A and 0B strings in the presence of
fluxes. These expressions are richer than the corresponding ones for
the bosonic noncritical string, both because of the flux dependence
and because they are nonperturbative in $\mu$. Our work should
generalise quite straightforwardly, particularly to the Euclidean type
0B case, and the correlators so obtained will contain nonperturbative
information about the theory.

A detailed investigation into the physical questions that motivated
the present exercise, namely a better understanding of the 2d black
hole background as well as of T-duality in the matrix model, is left
for subsequent work. We also note that the physical origin of the
Normal Matrix Model has not yet been understood. As it is clearly a
correct and useful description of the $c=1$ string, and moreover makes
sense only in the Euclidean context, it would be worth trying to put
it on a similar footing as MQM in terms of the dynamics of some
appropriate (Euclidean) D-branes.

\section*{Acknowledgements}

We are grateful to Shiraz Minwalla and Ari Pakman for useful
conversations, and to the people of India for generously supporting
our research. The research of AM was supported in part by CSIR Award
No. 9/9/256(SPM-5)/2K2/EMR-I.

\appendix
\section{Computation of two-point functions in the NMM}

Here we present some of the details of how to compute two-point
functions in the Normal Matrix Model. 
To start with, for the partition function we have
\bea
\label{zn2}
~
\nn {\cal Z}_{NMM}^{N=2}(t=0) &=& \int d^2z_1 d^2z_2\,|z_1-z_2|^2 \\
~
 && \times e^{-\n \left((z_1\bar{z}_1)^R + 
(z_2\bar{z}_2)^R\right) + \left(R\n-2 +
\frac{R-1}{2}\right)({\rm\,log}\,z_1\bar{z}_1 + {\rm\,log}\,z_2\bar{z}_2)}
~
\eea
As before, we change variables $z_i=\sqrt{m_i}\,e^{i\te_i}$, $d^2z_i\ra
dm_i\,d\te_i$ and we get
\bea
\label{zn2full}
~
\nn && {\cal Z}_{NMM}^{N=2}(t=0) = \int_0^\infty dm_1 dm_2 \int_0^{2\pi}
d\te_1 d\te_2\,\left(m_1 + m_2 - \sqrt{m_1 m_2}(e^{i\te_{12}} +
e^{-i\te_{12}})\right) \\
~
\nn && \hspace{2.6 cm} \times e^{-\n(m_1^R + m_2^R) + \left(R\n-2 +
\frac{R-1}{2}\right)({\rm\,log}\,m_1 + {\rm\,log}\,m_2)} \\ 
~
\nn &&= 4\pi^2 \int_0^\infty dm_1 dm_2\,(m_1 + m _2) (m_1 m_2)^{\left(R\n-2 +
\frac{R-1}{2}\right)} e^{-\n\left(m_1^R + m_2^R\right)} \\
~
&&= \sfrac{8\pi^2}{R^2} \nu^{-\left(\n + \half -
\frac{1}{2R}\right)}\G\left(\n + \shalf - \sfrac{1}{2R}\right) \times
\nu^{-\left(\n + \shalf - \sfrac{3}{2R}\right)}\G\left(\n + \shalf -
\sfrac{3}{2R}\right),
~
\eea
where $\te_{12}\equiv\te_1-\te_2$. In a similar manner we have
\bea
~
\nn && \del_{-2}\del_2 {\cal Z}_{NMM}^{N=2}(t=0) =  \int d^2z_1
d^2z_2\,|z_1-z_2|^2 (z_1^2 + z_2^2)(\bar{z}_1^2 + \bar{z}_2^2) \\
~
\nn && \hspace{3.5 cm}\times e^{-\n \left((z_1\bar{z}_1)^R +
(z_2\bar{z}_2)^R\right) + \left(R\n-2 +
\frac{R-1}{2}\right)({\rm\,log}\,z_1\bar{z}_1 + {\rm\,log}\,z_2\bar{z}_2)} \\
~
\nn &&= \int_0^\infty dm_1 dm_2 \int_0^{2\pi} d\te_1
d\te_2\,\left(m_1 + m_2 - \sqrt{m_1 m_2}(e^{i\te_{12}} +
e^{-i\te_{12}})\right) \\
~
\nn && \times \left(m_1^2 + m_2^2 + m_1m_2 e^{2i\te_{12}} + m_1m_2
e^{-2i\te_{12}}\right) e^{-\n(m_1^R + m_2^R) + \left(R\n-2 +
\frac{R-1}{2}\right)({\rm\,log}\,m_1 + {\rm\,log}\,m_2)} \\ 
~
\nn &&= 4\pi^2  \int_0^\infty dm_1 dm_2\,(m_1 + m_2)(m_1^2 + m_2^2) (m_1
m_2)^{\left(R\n-2 + \frac{R-1}{2}\right)} e^{-\n\left(m_1^R + m_2^R\right)}
~
\eea
Evaluating the integrals on $m_1, m_2$ we get
\bea
\label{dm2d2zn2}
\nn \del_{-2}\del_2 {\cal Z}_{NMM}^{N=2}(t=0) = 
~
\nn \sfrac{8\pi^2}{R^2} \nu^{-\left(\n + \half +
\frac{3}{2R}\right)}\G\left(\n + \shalf + \sfrac{3}{2R}\right) \times
\nu^{-\left(\n + \half - \frac{3}{2R}\right)}\G\left(\n + \shalf -
\sfrac{3}{2R}\right) \\
~
\nn + \sfrac{8\pi^2}{R^2} \nu^{-\left(\n + \half +
\frac{1}{2R}\right)}\G\left(\n + \shalf + \sfrac{1}{2R}\right) \times
\nu^{-\left(\n + \half - \frac{1}{2R}\right)}\G\left(\n + \shalf -
\sfrac{1}{2R}\right) \\
~
\eea
{}From \eref{zn2full} and \eref{dm2d2zn2} we have
\beq
\langle T_{-2/R} T_{2/R}
\rangle_{NMM}^{N=2}=\n^{-2/R}\left(\frac{\G\left(\n + \half +
\frac{3}{2R}\right)}{\G\left(\n + \half - \frac{1}{2R}\right)} +
\frac{\G\left(\n + \half + \frac{1}{2R}\right)}{\G\left(\n + \half -
\frac{3}{2R}\right)}\right)
\eeq
As before, to get the correct two point function we have to analytically
continue $\n=-i\m$ and take the modulus of the above expression. This
gives:
\beq
\label{ptm2t2}
\langle T_{-2/R} T_{2/R}\rangle_{NMM}^{N=2}=\mu^{-2/R}
\left|\frac{\G\left(\half - i\m +
\frac{3}{2R}\right)}{\G\left(\half - i\m - \frac{1}{2R}\right)} +
\frac{\G\left(\half - i\m + \frac{1}{2R}\right)}{\G\left(\half - i\m -
\frac{3}{2R}\right)}\right|
\eeq

\section{Evaluation of a summation in the MQM four-point function}
\label{evalsum}

In order to show the equivalence between Eqs. (\ref{ptm1t13}) and
(\ref{ptm1t13-summed}) we need to prove the following identity:
\beq
\label{sumid}
\sum_{n=0}^\infty
\frac{(-1)^n}{n!}\left(\frac{\G(-\frac{1}{R}+n)}{\G(-\frac{1}{R})}\right)^2
\frac{\G\left(\half - i\m + \frac{3}{2R} -n\right)}{\G\left( \half - i\m -
\frac{1}{2R}\right)} = \left(\frac{\G\left(\half - i\m +
\frac{1}{2R}\right)}{\G\left( \half - i\m - \frac{1}{2R}\right)}\right)^2
\eeq

\nt Let us start with the expression:
\beq
\label{expr}
\cE = \sum_{n=0}^\infty
\frac{(-1)^n}{n!}\left(\frac{\G(-\sfrac{1}{R}+n)}{\G(-\sfrac{1}{R})}\right)^2
\G\left(\shalf - i\m + \sfrac{3}{2R} -n\right)
\eeq
Using the integral representation of the $\G$ function we write this as:
\beq
\cE = \frac{1}{\left(\G(-\frac{1}{R})\right)^2}\sum_{n=0}^\infty
\frac{(-1)^n}{n!}\int d^3
t~(t_1t_2)^{-\frac{1}{R}+n-1}t_3^{-n+\half-i\m+\frac{3}{2R}-1}e^{-t_1-t_2-t_3}
\eeq
The sum over $n$ can now be performed immediately and we have:
\bea
~
\nn \cE &=& \frac{1}{\left(\G(-\frac{1}{R})\right)^2} \int d^3
t~e^{-\frac{t_1t_2}{t_3}}
(t_1t_2)^{-\frac{1}{R}-1}\,t_3^{\half-i\m+\frac{3}{2R}-1}\,e^{-t_1-t_2-t_3} \\
~
&=& \frac{1}{\left(\G(-\frac{1}{R})\right)^2} \int d^3
t~e^{-t_1(1+\frac{t_2}{t_3})}
(t_1t_2)^{-\frac{1}{R}-1}\,t_3^{\half-i\m+\frac{3}{2R}-1}\,e^{-t_2-t_3}
~
\eea
Using the change of variables $t_1\to t_1(1+\frac{t_2}{t_3})$ and performing
the integral on $t_1$ we get:
\beq
\cE = \frac{1}{\G(-\frac{1}{R})} \int d^2 t~
t_2^{-\frac{1}{R}-1}\,t_3^{\half-i\m+\frac{3}{2R}-1}
\,t_3^{-\frac{1}{R}}\,(t_2+t_3)^\frac{1}{R}\,e^{-t_2-t_3}
\eeq
We next introduce a parameter $\al$ which allows us to write the above
equation as:
\beq
\cE = \frac{1}{\G(-\frac{1}{R})}
\left(-\frac{\del}{\del\al}\right)^\frac{1}{R}\left.\int d^2 t~
t_2^{-\frac{1}{R}-1}\,t_3^{\half-i\m+\frac{1}{2R}-1}\,
e^{-\al(t_2+t_3)}\right|_{\al=1}
\eeq
Changing variables $t_i\to\al t_i$ we have:
\bea
~
\nn \cE &=& \frac{1}{\G(-\frac{1}{R})}
\left(-\frac{\del}{\del\al}\right)^\frac{1}{R}\left.\al^{-\half+i\m+
\frac{1}{2R}}\right|_{\al=1}\int d t_2~ t_2^{-\frac{1}{R}-1}
\,e^{-t_2}\int d
t_3~t_3^{\half-i\m+\frac{1}{2R}-1}\,e^{-t_3} \\
~
&=& \G(\shalf-i\m+\sfrac{1}{2R})\left(-\frac{\del}{\del\al}\right)
^\frac{1}{R}\left.\al^{-\half+i\m+\frac{1}{2R}}\right|_{\al=1}
~
\eea
Using the relation:
\beq
\left(-\frac{\del}{\del\al}\right)^m\al^n\Big|_{\al=1} =
\frac{\G(-n+m)}{\G(-n)}
\eeq
we finally have:
\beq
\cE =
\frac{\left(\G(\half-i\m+
\frac{1}{2R})\right)^2}{\G(\half-i\m-\frac{1}{2R})}
\eeq
Using \eref{expr} and dividing both sides by
$\G(\half-i\m-\frac{1}{2R})$ we immediately get \eref{sumid}.

\section{Four-point function in NMM}
\label{fourpoint}

We now briefly describe the calculation of the connected four-point
function of unit momentum modes in the NMM. This is obtained by
differentiating the free energy $\cF$ with respect to the
couplings. We thus have:
\bea
\nn \langle (T_{-1/R}T_{1/R})^2\rangle &=& \del_{-1}^2\del_1^2 \cF \\
&=&  \langle(T_{-1/R}T_{1/R})^2\rangle^{\rm disconn} 
-2\langle T_{-1/R}T_{1/R}\rangle^2
\eea
where $\cF=\ln \cZ_{NMM}$. The second term in the above equation
can be calculated from the NMM with $N=2$ and is given by:
\bea
\nn \langle (T_{-1/R}T_{1/R})^2\rangle^{\rm disconn} &=& \langle
(\tr Z^\dag)^2 (\tr Z)^2\rangle_{NMM}^{N=2}\\
\nn &=& \n^{-2/R}\left(\frac{\G\left(\n + \half +
\frac{3}{2R}\right)}{\G\left(\n + \half - \frac{1}{2R}\right)} +
\frac{\G\left(\n + \half + \frac{1}{2R}\right)}{\G\left(\n + \half -
\frac{3}{2R}\right)}\right)
\eea
The explicit calculation is very similar to the calculation of
$\langle T_{-2/R}T_{2/R}\rangle$ from the NMM. The disconnected piece
is simply the square of the two-point function listed in
\eref{2point}.  Putting everything together the connected four-point
function is given by:
\bea
\nn\langle (T_{-1/R}T_{1/R})^2\rangle^{\rm conn}_{NMM} 
&=& \n^{-2/R}\left[\frac{\G\left(\n + \half +
\frac{3}{2R}\right)}{\G\left(\n + \half - \frac{1}{2R}\right)} +
\frac{\G\left(\n + \half + \frac{1}{2R}\right)}{\G\left(\n + \half -
\frac{3}{2R}\right)}\right.\\ && 
~~~~\left. - 2 \left(\frac{\G(n+\half + \frac1{2R})}
{\G(n+\half + \frac1{2R})}\right)^2\right]
\eea

Analytically continuing $\n=-i\mu$ and taking the modulus, and then
using the definition of $\cF^+$ in \eref{functionf}, we finally get:
\beq
\langle (T_{-1/R}T_{1/R})^2\rangle^{\rm conn}_{NMM} =
(\mu)^{-2/R}
\left|\cF^+(\sfrac{3}{2R},-\sfrac{1}{2R};\mu)
-\shalf \Big(\cF^+(\sfrac{1}{2R}, - \sfrac{1}{2R};\mu)\Big)^2
\right|
\eeq

\section{$2n$-point functions in NMM}
\label{2npointfunctions}

Here we present the detailed calculation of the $2n$-point
functions from the NMM. In what follows we will take the rank of the
matrix, $N$, to be equal to $n$. We have:
\bea
&& \nn (\del_{-1}\del_1)^n {\cal Z}_{NMM}^{N=n}(t=0) = 
\int\prod_{i=1}^n d^2z_i
\prod_{i<j}^n |z_i-z_j|^2 \left(\sum_{i=1}^n z_i\right)^n 
\left(\sum_{i=1}^n \zbar_i\right)^n \\ 
~
&& \hspace{4.5cm} \times e^{-\n \sum_{r=1}^n (z_r\zbar_r)^R
+ \left(R\n-n +
\frac{R-1}{2}\right)\sum_{r=1}^n{\rm\,log}\,z_r\zbar_r} 
\eea
We would now like to make the substitution
$z_i=\sqrt{m_i}\,e^{i\theta_i}$ and perform the $\te$ integrals. The
remaining integrand will then be a function of the $m_i$ and we will
find that it has the form $\left(\sum_{\{k_i\}} C(\{k_i\})^2\prod_i
m_i^{k_i} + {\rm permutations}\right)e^{-S_{NMM}}$. Here $\{k_i\}$ are
positive integers corresponding to {\it strictly ordered} partitions
of $n(n+1)/2$, i.e.:
\beq
\sum_{i=1}^n k_i = \sfrac{n(n+1)}{2},\quad
k_1>k_2>\cdots> k_n\ge 0
\eeq
The permutations referred to are
of the $m_i$. Because the $m_i$ are integration variables, summing
over permutations simply amounts to multiplying by a factor of
$n!$. The constant coefficients have been labelled $C(\{k_i\})^2$ in
anticipation of the fact that they will turn out to be squares. After
performing the integration over $m_i$ and dividing by $Z_{NMM}$ we get
the final answer as a sum of ratios of products of gamma functions,
with each term in the sum corresponding to a strictly ordered 
partition $\{k_i\}$ of $n(n+1)/2$.

We will first show that the coefficients are perfect squares
$C(\{k_i\})^2$. After that we will turn to the calculation of the
$C(\{k_i\})$. Consider the expression:
\beq
\label{texpr}
\cU = \left(\sum_{i=1}^n z_i\right)^n
\prod_{j<k}^{n}(z_j-z_k).
\eeq
The full integrand is then $\cU\bar\cU$ times the exponential factor.
Because the action is independent of the $\theta$'s, the entire
$\te$-dependence of the integrand is in $\cU\bar\cU$. Note that $\cU$
has only positive powers of $e^{i\theta_i}$ and $\bar\cU$ has only
negative powers. Only the $\te$-independent terms in the expansion of
$\cU\bar\cU$ will survive the $\te$ integrals. 

It is easy to see that if we expand $\cU,\bar\cU$ then we get:
\bea
\label{ttbar}
~
\nn\cU &=& \sum_{\{\al_i\}}
C(\{k_i\})\prod_{i=1}^n z_i^{k_i}
+ {\rm permutations} \\
~
\bar\cU &=& \sum_{\{\al_i\}}
C(\{k_i\})\prod_{i=1}^n \zbar_i^{k_i}
+ {\rm permutations},
~
\eea
with $\{k_i\}$ defined as before. It is now clear that the
coefficients of $\te$-independent terms in $\cT\bar\cT$ must be
perfect squares, as the phase of a term in the first expression
of \eref{ttbar} can only be cancelled by the complex conjugate term
from the second expression, which has the same coefficient as the
first term.

Let us now determine the coefficients $C(\{\al_i\})$. First we
note the following property of the positive phase part of the
Vandermonde:
\beq
\label{vperm}
\prod_{j<k}^{n}(z_j-z_k)
=\sum_\cP (-1)^\cP \prod_{j=1}^n
z_j^{\cP_j},
\eeq
where $\cP$ is a particular permutation of the $n$ integers
($n-1,n-2,\cdots,0$) and $\cP_j$ denotes the $j^{\rm th}$ element of the
permutation $\cP$\footnote{For example, $\cP_j=n-j$ when $\cP$ is the 
identity permutation.}. The sign for the first permutation is positive by
construction. Any other permutation can be arrived at by a series of
interchanges $z_i\leftrightarrow z_j$. Each such interchange
introduces a minus sign in the Vandermonde. Thus even permutations
have a positive sign, while odd permutations have a negative sign,
leading to \eref{vperm}. Expanding the first factor in \eref{texpr}
in a multinomial series and using \eref{vperm} we get:
\bea
~
\nn \cU &=& \left(\sum_{\{\be_i\}}\prod_{i=1}^n\ncr{n-\sum_{j=1}^{i-1}
\be_j}{\be_i} z_i^{\be_i}\right)
\left(\sum_\cP (-1)^\cP \prod_{j=1}^n
z_j^{\cP_j}\right) \\
~
&=& \sum_{\{\be_i\}}\sum_\cP (-1)^\cP \prod_{i=1}^n\ncr{n-\sum_{j=1}^{i-1}
\be_j}{\be_i}z_i^{\be_i+\cP_i}
~
\eea
where $\{\be_i\}$ are the {\it unordered} partitions of $n$.

Let us examine the possible values of the exponent $k_i=\beta_i + \cP_i$
in the above. If $k_i=k_j$ for some $i\ne j$
then the corresponding coefficient is zero. This can be traced back to
the fact that the expression \eref{texpr} is odd under pairwise
interchange of the $z$'s. Therefore we can rewrite the above as:
\bea
\cU = \sum_{\stackrel{\scriptstyle k_i\ne k_j}{\sum_i k_i =
n(n+1)/2}}\sum_{\cP}(-1)^\cP 
\prod_{i=1}^n \ncr{n-\sum_{j=1}^{i-1}(k_j-\cP_j)}{k_i-\cP_i}
\prod_{i=1}^n z_i^{k_i}
\eea
Because the $k_i$ are all distinct, we can limit ourselves to strictly
ordered sets satisfying $k_1>k_2>\cdots k_n$. The other orderings are
obtained by permuting these ones, or equivalently by permuting the
$z_i$'s. Thus we have:
\beq
\cU = \sum_{\stackrel{\scriptstyle k_1>k_2>\cdots >k_n}{\sum_i k_i =
n(n+1)/2}} C(\{k_i\}) \prod_{i=1}^n z_i^{k_i} + \hbox{(permutations of }z_i)
\eeq
with 
\beq
\label{ck}
C(\{k_i\}) = 
\sum_{\cP}(-1)^\cP 
\prod_{i=1}^n \ncr{n-\sum_{j=1}^{i-1}(k_j-\cP_j)}{k_i-\cP_i}
\eeq
Finally we combine $\cU$ with $\bar\cU$ and integrate over the angles
to get:
\bea
&&\nn (\del_{-1}\del_1)^n {\cal Z}_{NMM}^{N=n} = 
(2\pi)^n\int\prod_{i=1}^n dm_i \sum_{\{k_i\}} C(\{k_i\})^2 
\prod_{i=1}^n m_i^{k_i}
e^{\sum_{i=1}^n (-\n m_i^R + (R\n-n + \frac{R-1}{2}){\rm\,log}\,m_i)}\\
&&\hspace{7cm} + {\rm permutations}\\
&&\nn\hspace{1cm} = (2\pi)^n n!\sum_{\{k_i\}} C(\{k_i\})^2\prod_{i=1}^n 
\n^{-\left(\half+\n +(k_i-n+\half)\frac{1}{R}\right)}
\G\left(\shalf+\n +(k_i-n+\shalf)\sfrac{1}{R}\right)  
\eea
Using the expression for the partition function $\cZ_{NMM}$ from
\eref{znnfull2} for $N=n$ we have:
\beq
\frac{(\del_{-1}\del_1)^n
\cZ_{NMM}^{N=n}}{\cZ_{NMM}^{N=n}}
= \n^{-n/R}\,\sum_{\{k_i\}} C(\{k_i\})^2\prod_{i=1}^n 
\frac{\G\left(\shalf+\n +(k_i-n+\shalf)\sfrac{1}{R}\right)}
{\G\left(\shalf+\n -(i-\shalf)\sfrac{1}{R}\right)}
\eeq
The $2n$-point function is given by analytically continuing $\n=-i\m$,
changing to MQM normalisation using \eref{nmmnorm} (which amounts to
removing the power of $\n$ in front), and finally taking the modulus:
\beq
\langle(T_{-1/R}T_{1/R})^n\rangle = \left|\sum_{\{k_i\}}
C(\{k_i\})^2\prod_{i=1}^n \frac{\G\left(\shalf-i\m
+(k_i-n+\shalf)\sfrac{1}{R}\right)} {\G\left(\shalf-i\m
-(i-\shalf)\sfrac{1}{R}\right)}\right|
\eeq
with $C(\{k_i\})$ given by \eref{ck}.

\end{document}